\documentclass[aps,pra,preprintnumbers,amsmath,amssymb,floatfix,twocolumn]{revtex4-2}
\usepackage{graphicx}
\usepackage{indentfirst}
\usepackage{color,soul}
\usepackage{braket}
\usepackage{tabularx}
\usepackage{hyperref}
\usepackage{mathtools}

\begin{document}

\title{Quantum-secured single-pixel imaging with enhanced security}

\author{Jaesung Heo}
\author{Junghyun Kim}
\author{Taek Jeong}
\author{Yong Sup Ihn}
\author{Duk Y. Kim}
\author{Zaeill Kim}
\author{Yonggi Jo}\email{yonggi@add.re.kr}
\affiliation{Advanced Defense Science \& Technology Research Institute, Agency for Defense Development, Daejeon 34186, South Korea}

\date{\today}

\begin{abstract}
In this paper, we propose a novel quantum-secured single-pixel imaging method that utilizes non-classical correlations of a photon pair. Our method can detect any attempts to deceive it by exploiting a non-classical correlation of photon pairs while rejecting strong chaotic light illumination through photon heralding. A security analysis based on polarization-correlation has been conducted, demonstrating that our method has improved security compared to existing quantum-secured imaging. More specifically, a partial deceiving attack, which sends a mixture of a true and a false signal, can be detected with our proposed analysis, while currently employed methods cannot. We also provide proof-of-principle demonstrations of our method and trustworthy images reconstructed using our security analysis. Our method can be developed using matured techniques used in quantum secure communication, thus offering a promising direction for practical applications in secure imaging.
\end{abstract}

\maketitle

\section{Introduction}

Non-classical correlations in quantum systems are fundamental sources of quantum advantage in various quantum information protocols. Entanglement-based quantum key distribution (QKD) \cite{E91,BBM92} and quantum-secured imaging \cite{Malik2012,Malik2012IEEE} utilize quantum correlations to provide security against potential eavesdropping attacks in quantum channels, while quantum ghost imaging (QGI) \cite{Pittman1995} employs the correlations to enhance the signal-to-noise ratio (SNR) of an image beyond the classical limit.

Active imaging systems can be threatened by a deceiving attack, i.e., the imaging system constructs a false image based on fake signals sent from an adversary. For example, there can be an adversary who tries to fool an imaging system into showing a bird instead of an aircraft to avoid detection. For detecting these attempts to deceive the system, quantum-secured imaging (QSI) has been proposed \cite{Malik2012}, which was based on a prepare-and-measure approach. In this method, an attack can be detected by analyzing an error of polarization information between a photon sent and a photon received. If there is an adversary, then the error exceeds a security threshold, and the obtained image from the system is totally untrustworthy.

A prepare-and-measure approach needs a perfect single photon generator for realization of theoretical security. However, since there is no such device under current technologies, a weak coherent pulse has been used for QKD. A weak coherent pulse can carry two or more photons since it follows the Poisson distribution. This multi-photon pulse can degrade security. In QKD, a decoy state method is employed to solve this security issue \cite{Hwang2003}. Conversely, for prepare-and-measure QSI, applying a decoy state method is not trivial; thus, its security can be threatened under current technologies. Instead of prepare-and-measure approaches, we can use a correlation-based protocol. In correlation-based QSI \cite{Malik2012IEEE}, non-classical correlations can be analyzed to detect any attempts to deceive the imaging system, and the SNR of an image can be enhanced similarly to QGI. Moreover, a correlation-based protocol can provide better loss tolerance in a free-space channel compared to a prepare-and-measure approach due to the suppression of dark count effects in a single-photon detector \cite{Waks2002}. Recently, an experimental demonstration of quantum-secured ghost imaging in the time-frequency domain has been reported \cite{Yao2018}.

In a prepare-and-measure QSI, the same photon is used for both imaging and security check. However, in existing correlation-based QSI \cite{Malik2012IEEE, Yao2018}, the imaging and security check are carried out sequentially using different photon pairs. Furthermore, while the security analysis of the existing proposal focuses on the most rudimentary scenario, another potential attack against the system would remain.

In this article, we propose a novel quantum-secured single-pixel imaging (QS-SPI) method that exploits non-classical correlations of a photon pair for imaging and security checking, simultaneously. Our imaging method is based on single-pixel imaging (SPI), also known as computational ghost imaging (CGI). Compared to existing QSI methods, our proposed method offers enhanced security against potential attacks. Specifically, we consider a deceiving attack that constructs a fake target image by intercepting genuine signals and illuminating false signals. By analyzing non-classical correlations of photon pairs, the attack can be exposed by an error rate. However, existing methods are unable to detect a partial deceiving attack, which creates a deceiving image with an error rate below the detection threshold by mixing true and false signals. QS-SPI can detect this attack by analyzing not only the non-classical correlations of photon pairs, but also the spatial profiles of the obtained images. Furthermore, our method can extract trustworthy information while discarding delusive information under a partial deceiving attack. As QS-SPI conducts imaging and security checking simultaneously, the obtained images are closely related to the security of the protocol, thereby enhancing the security. We demonstrate our theory through proof-of-principle experiments based on polarization-correlation. We expect that advanced techniques used in quantum secure communication can further improve the security of QS-SPI.

This article is organized as follows. In Sec. \ref{Sec2}, we propose the QS-SPI setup and describe its security. The experimental realization of our method is shown in Sec. \ref{Sec3}, and we conclude our study in Sec. \ref{Sec4}.

\section{Quantum-Secured Single-Pixel Imaging}\label{Sec2}

\begin{figure}[t!]
	\centering\includegraphics[width=0.4\textwidth]{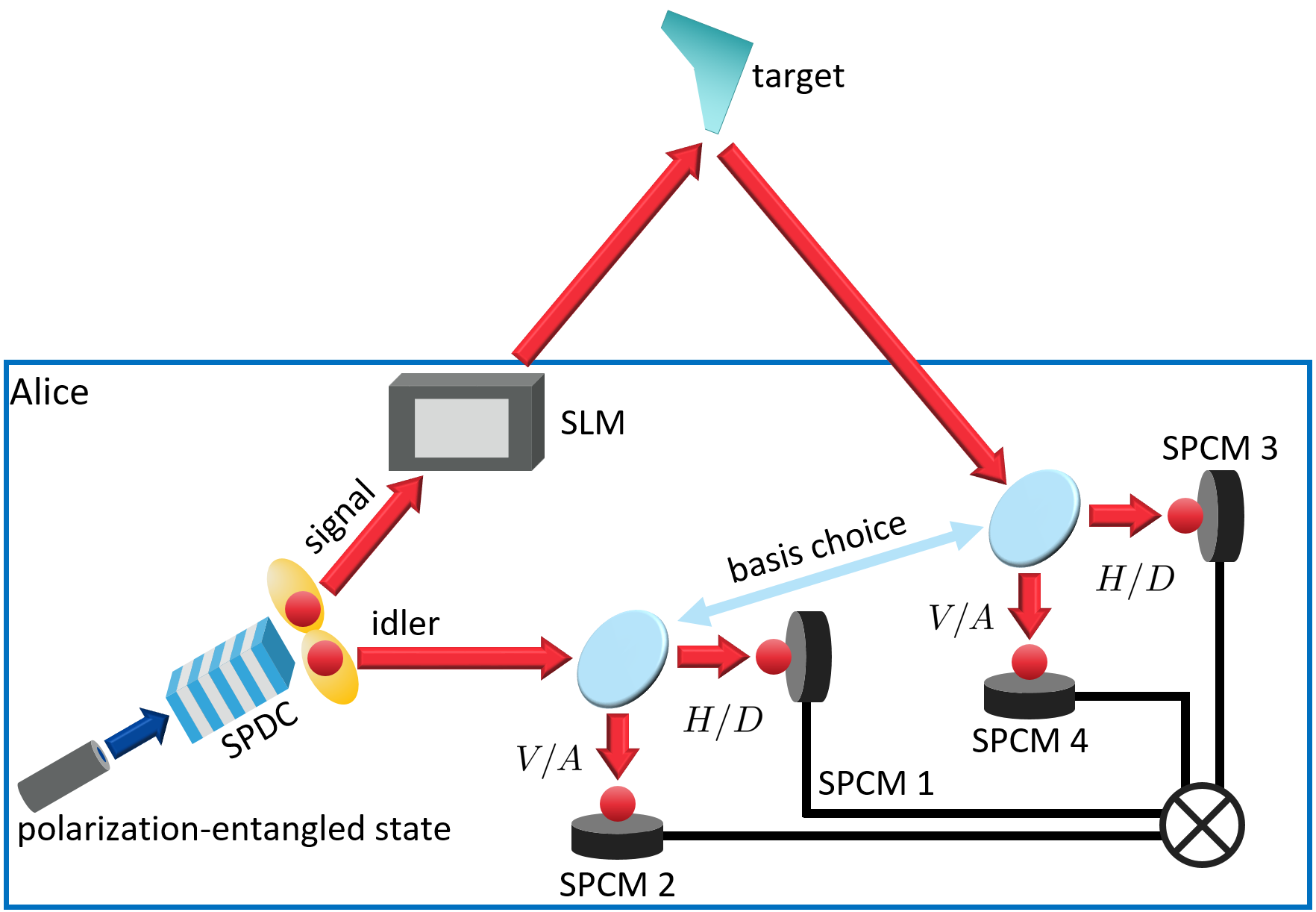}%
	\caption{A schematic diagram of QS-SPI. Alice, who operates the imaging system, generates pairs of polarization-entangled photons. The signal photon is sent to an SLM to be filtered according to a desired imaging pattern. This filtered photon illuminates the target and is reflected to be measured by SPCMs. The idler photon, the other photon in the entangled pair, is measured directly by the other set of SPCMs. The time-correlation and polarization-correlation of the two modes are analyzed from the measured data of the SPCMs.}\label{Scheme}
\end{figure}

In SPI, an image is constructed using the correlation between the spatial information of a beam, given by a spatial light modulator (SLM), and the intensity of that beam after interacting with a target. Single-pixel detectors, such as photodiodes, are used to measure the intensity. Let us denote the $k\text{-th}$ spatial pattern on the SLM as $P^{(k)}$ and its corresponding intensity measured as $I^{(k)}$. Varying the spatial patterns, the corresponding intensities are recorded, and the correlation of the two yields a target image $G$,
\begin{align}\label{GI}
	G(i,j)=\braket{P^{(k)}(i,j)I^{(k)}}-\braket{P^{(k)}(i,j)}\braket{I^{(k)}},
\end{align}
where $i$ and $j$ represent the pixel position of the 2D image and $\braket{\cdot}$ denotes the average of all of the $N$ patterns \cite{Shapiro2008, Gibson2020}.

Various imaging patterns can be used for $P^{(k)}$, but to enhance the image quality while reducing the number of patterns required for imaging, an orthogonal pattern set is selected. One type of orthogonal pattern set used for SPI is Hadamard patterns \cite{Pratt1969, Souza1988, Duarte2008, Sun2017}, which are constructed based on the Hadamard matrix. A $2^{n+1}\times 2^{n+1}$ Hadamard matrix is calculated by the following:
\begin{align}
	H_{2^{n+1}}=H_{2^{n}}\otimes H_{2},
\end{align}
where
\begin{align}
	H_{2}=\begin{pmatrix}
		1 & 1 \\
		1 & -1
	\end{pmatrix},
\end{align}
and $\otimes$ denotes the tensor product. Hadamard patterns are generated by reshaping each row of the Hadamard matrix $H_{2^{2n}}$ into a $2^{n}\times 2^{n}$ square matrix. Since the negative pixel value cannot be displayed in a digital-micromirror-device (DMD), we exploit two shots to represent a Hadamard pattern in our demonstration \cite{Gibson2020}. However, we expect this to be reduced to one-shot detection with various techniques for practical implementation \cite{Yu2021}. Transitioning matrix elements of 1 as white and -1 as black, one imaging pattern is made, and the other one is the inverse. For a $2^n \times 2^n$ resolution image, the total number of shots required for imaging is $2^{2n+1}$.

Fig.~\ref{Scheme} shows a schematic diagram of QS-SPI. Polarization entangled photon pairs generated via spontaneous parametric down-conversion (SPDC) are exploited for the security check. The Bell state used is $\ket{\Phi^{+}}=\frac{1}{\sqrt{2}}\left(\ket{H,H}_{SI}+\ket{V,V}_{SI}\right)$, where $\ket{\cdot}$ represents the polarization state of a single photon, $H$ ($V$) denotes horizontal (vertical) polarization, and the subscripts $S$ and $I$ denote signal and idler modes, respectively. The security of QS-SPI is based on the time and polarization correlation of photon pairs.

The signal photon is then sent to an SLM. Based on an imaging pattern, the SLM filters signal spatially, and the selected photon illuminates a target. The idler photon is ideally retained and measured to analyze correlations with the signal photon.

In the QS-SPI setup, four single photon counting modules (SPCMs) are exploited as single-pixel detectors, which have a sub-ns timing resolution \cite{Ndagano2020,Li2021,Defienne2021}. They provide a simpler measurement setup compared to a QGI setup exploiting electron multiplying charge-coupled device (EMCCD) \cite{Gregory2020, Gregory2021}. As SPCM has a higher acceptable noise level than EMCCD, QS-SPI can reconstruct a trustworthy image even if an intensity of false signal reaches the EMCCD saturation level. Moreover, given the timing resolution of SPCMs, we can exploit a time-correlation of photon pairs to reject a strong background noise \cite{Yang2020,Kim2021APL,Johnson2022}, while a strong chaotic light can severely disturb a conventional CGI system \cite{Li2021,Kim2022,Heo2022}. QS-SPI uses coincidence counts of the signal and idler modes as the imaging intensity $I$ to exploit time-correlation. Thus, it is naturally immune to an imaging disrupting attack: an attack where strong chaotic light illuminates an imaging system to saturate the sensor.

Note that it is suitable to use a DMD as an SLM of QS-SPI rather than liquid crystal based SLM. To control and measure the correlations of signals at the single-photon level, an SLM with high reflectivity and polarization-independence is required. Compared to a liquid crystal based SLM, a DMD is suitable for such purposes.

\subsection{Method of image deceiving attack}

The security analysis method of quantum key distribution (QKD) has been well-established for protecting photon-carried information against eavesdropping, which is directly related to the generation of secret keys. For example, in polarization-based QKD, the polarization-encoded information of a photon is critical to the secret key. There are many advanced attacks for extracting polarization-encoded information of successively transmitted photons, such as collective attacks which exploit demanding technologies including quantum cloning machines \cite{Buzek1996, BrussEkert1998}, quantum memories \cite{Lvovsky2009}, and collective measurements \cite{Biham1997, Acin2007}. However, in SPI, the main purpose of an attack is to deceive an imaging system to construct a fake image rather than to eavesdrop on secret keys. For this purpose, a meaningful attack is to modulate the intensity (photon rate) of the light for the formulation of a fake image. Under this circumstance, an intercept-and-resend attack is expected as the optimal attack strategy for image-deceiving attacks \cite{Malik2012}.

\begin{figure}[t!]
	\centering\includegraphics[width=0.4\textwidth]{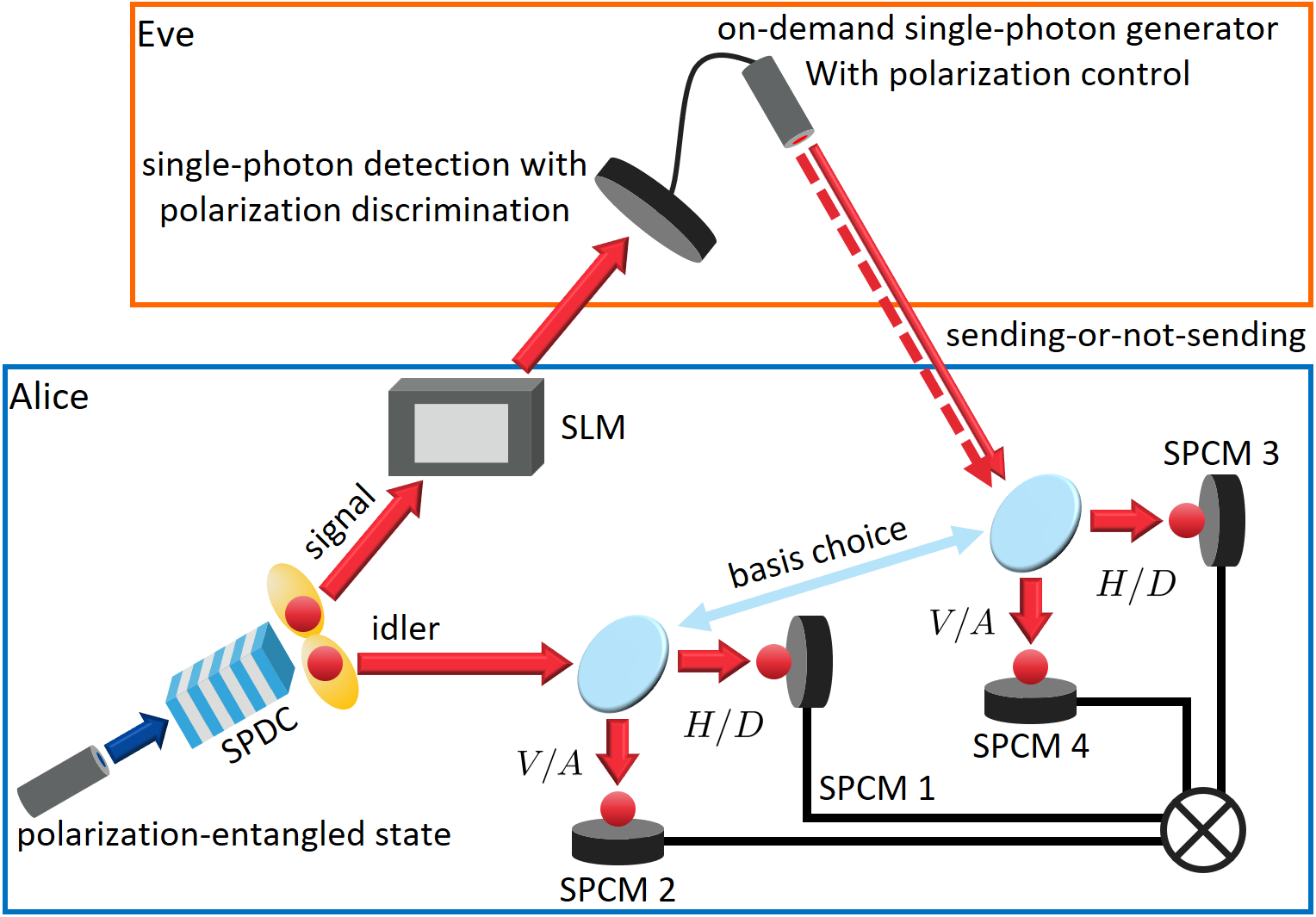}%
	\caption{A schematic diagram of QS-SPI under a potential attack by an adversary, referred to as Eve. In this attack, Eve manipulates the photon rate of a fake signal for the purpose of generating a fraudulent image, which is then sent to Alice. To evade detection, Eve must ensure that the polarization of the fake signal is similar to Alice's signal, which can be achieved by using an intercept-and-resend attack.}\label{SchemeEve}
\end{figure}

Fig.~\ref{SchemeEve} shows a schematic diagram of QS-SPI under a possible attack of an enemy called Eve. It is assumed that Eve can exploit all implementations allowed by the laws of physics and all processes of QS-SPI are known to Eve. For the deceiving attack, Eve possesses time-resolved single-photon detectors with polarization discrimination and an on-demand single-photon source with polarization control. Eve intercepts Alice's signal photon and discriminates its polarization. Since it is not possible to simultaneously measure a quantum state in the conjugate bases, disturbance of the original photon state is inevitable. 
After the polarization measurement, the on-demand single-photon source generates a photon in the measured polarization, and the photon is sent to Alice. SPI constructs an image by spatial pattern information and received photon rate; therefore, Eve should control ${n_g}/{n_m}$ according to the DMD pattern to make QS-SPI construct a fraud image, where $n_g$ ($n_m$) is a generated (measured) photon rate of Eve.

As the signal and the idler are a polarization entangled photon pair, polarization of the signal is heralded when polarization of the idler is measured. However, an intercept-and-resend attack leads to detection of the signal photons in an unheralded polarization. Such errors are key resources for security analysis of QS-SPI. Details of the security check are described in the following section.

\subsection{Security analysis in QS-SPI}

The presence of Eve is tested by Alice via measuring photons in mutually unbiased bases (MUBs). One basis, named rectilinear basis, consists of horizontal and vertical polarization, and the other basis, diagonal basis, does diagonal ($D$) and anti-diagonal ($A$) polarization. For the two bases, the following relations are satisfied: $\ket{D}=\frac{1}{\sqrt{2}}\left(\ket{H}+\ket{V}\right)$ and $\ket{A}=\frac{1}{\sqrt{2}}\left(\ket{H}-\ket{V}\right)$; thus, the two bases are MUBs. Alice, who has a QS-SPI system, randomly chooses the measurement basis for the security check. Unlike QKD, it is not necessary for the basis choice of the signal and idler modes to be independently random since the measurement setups of both modes belong to Alice.

Let us define $r_{1}\coloneqq H$, $r_{2}\coloneqq V$, $d_{1}\coloneqq D$, and $d_{2}\coloneqq A$. Then, $P(X_{i},X_{j})=C(X_{i},X_{j})/\sum_{k,l=1}^{2}C(X_{k},X_{l})$ is satisfied, where $C(x,y)$ is the coincidence counts of $x$- and $y$-polarized photons in the signal and idler modes, respectively. $P(x,y)$ is the probability of a coincidence count $C(x,y)$ to happen, $X\in\{r,d\}$, and $i,j\in\{1,2\}$. From $\ket{\Phi^{+}}$, $P(X_{i},X_{i})=1/2$ and $P(X_{i},X_{j})=0$ for $i\neq j$, indicating that the latter coincidence count is erroneous. The error rate can be defined as the ratio of erroneous coincidence counts to all coincidence counts. Since the idler photon is unhindered by Eve's attack, an error rate is defined with respect to polarization of the idler. Thus, a polarization error rate of an $X_{i}$-polarized idler is calculated as, 
\begin{align}\label{polE}
	e_{X_{i}}&=\frac{C(X_{j},X_{i})}{\sum_{k=1}^{2}C(X_{k},X_{i})},
\end{align}
where $i\neq j$.
Under no attack, the error rates are always zero. However, if there is an enemy who tries to disturb the imaging system, erroneous coincidence counts increase; therefore, Alice can notice the presence of an attack by analyzing the error rates.

Eve possesses its own MUBs for polarization measurement. Let us denote its constitutive polarization in the primed notation, i.e., $H'$, $V'$, $D'$, and $A'$. For their respective identical basis choice, let the angle difference between the two as $\theta$, measured counterclockwise from one polarization of the Alice to that of the Eve, i.e., the angle measured from the $H$-polarization to $H'$-polarization, counterclockwise. Then, the angle difference between their respective different bases is $\theta\pm\frac{\pi}{4}$.

Suppose Alice chooses rectilinear basis and measures the idler in $H$-polarization. For Eve's own rectilinear basis, Eve measures Alice's signal in $H'$ ($V'$)-polarization with probability $\cos^2\theta$ ($\sin^2\theta$).
Regardless of Eve's result, Alice's error rate, i.e., the probability of detecting $V$-polarized signal, is $\cos^2\theta\sin^2\theta$. Thus, the error rate observed by Alice is $2\cos^2\theta\sin^2\theta=(1-\cos^2 2\theta)/2$. If Eve chooses the other measurement basis, the error rate is calculated by replacing $\theta$ to $\theta\pm\frac{\pi}{4}$: $(1-\sin^2 2\theta)/2$. Since Eve's basis choice is random, the error rate in the $H$-polarized idler is calculated as follows: $e_{H}=\left[(1-\cos^2 2\theta)/2+(1-\sin^2 2\theta)/2\right]/2=1/4$.
The result indicates that the threshold error rate for determining the presence of an attack is 25\% regardless of the angle difference $\theta$. If an error rate in either rectilinear or diagonal basis is greater than or equal to 25\%, the protocol is compromised.

Rather than a full deceiving attack, Eve can perform a partial attack, i.e., Eve performs the attack against some photons, while the other photons are passed through. With this attack, Eve cannot manipulate a full image, but can add some pixels in an original image. For example, when an original image is "1", Eve can make the image "4" by adding one horizontal and one diagonal line. Even when the attack is performed, the error rate can remain less than $25\%$, since only a part of photons influenced by Eve contributes to the error rate. Thus, the existing security analysis of QSI methods cannot detect the presence of Eve, even if the image is modified from the original image. To detect this partial deceiving attack, we should analyze the intensity of Eve's light and the pixel-area that Eve attempts to modify. Here, we provide a security analysis to show how the parameters can be used to detect the attack and how we can reconstruct a trustworthy image.

Let the coincidence rate originated from the signal of Alice (Eve) without target information be $n_{A(E)}$. By a target profile $\chi_{A(E)}$ and channel efficiency $\eta_{A(E)}$, detected coincidence counts $I_D$ at the $n$-th imaging pattern is
\begin{align}
    I_D^{(n)} &= \sum_{k\in\{A,E\}} \frac{1}{M}\sum_{i,j}P^{(n)}(i,j)\left( \eta_k(i,j) \chi_k(i,j) n_k \right) \\
    &\coloneqq \sum_{k\in\{A,E\}} I_k^{(n)},
\end{align}
where $M$ is the total number of pixels and $I_{A(E)}$ is detected coincidence counts originating only from Alice's (Eve's) signal. As the sum of any Hadamard pattern to its corresponding inverse pattern is the pattern where all components are 1, the total coincidence counts for the whole $N$ imaging patterns are
\begin{align}
    \sum_n I_D^{(n)}=\sum_{k\in\{A,E\}}\frac{N}{2M} S_k n_k,
\end{align}
where $S_{A(E)}$ is the sum of components of Alice's (Eve's) target image, i.e., $S_{A(E)} \coloneqq \sum_{i,j}\eta_{A(E)}(i,j)\chi_{A(E)}(i,j)$.
Previous analysis shows that the minimum erroneous coincidence counts are a quarter of $I_E$. Thus, the threshold error rate $e_T$ becomes
\begin{align}\label{pda errorT by counts}
    e_T=\frac{1}{4}\frac{S_E n_E}{S_A n_A + S_E n_E}.
\end{align}

It is explicitly shown that the error rate under a partial deceiving attack can be smaller than 25\%. Also, the error rate is closely related not only to the beam intensity but also to the target profile. However, one cannot get target information before imaging. Thus, to notice Eve's partial deceiving attack, a threshold error rate based on the analysis on constructed image is required.

From Eq.~\ref{GI}, an image formed by $I_k$, where $k\in\{A,E\}$, is
\begin{align}
    G_k(i,j)&=\frac{1}{N} \sum_n P^{(n)}(i,j)I_k^{(n)} - \frac{1}{2N}\sum_n I_k^{(n)} \\
    &=\frac{1}{4M} \eta_k(i,j) \chi_k(i,j) n_k.
\end{align}
Then, an image formed by $I_D$ is $G_\text{all}=G_A+G_E$. Note that neither $G_A$ nor $G_E$ are directly obtainable. An image formed by correct (erroneous) coincidence counts is obtainable, denoted as $G_{\text{cor(mask)}}$. Under an intercept-and-resend attack, $G_\text{mask} = \frac{1}{4} G_E$, which makes $G_\text{cor} = G_A + \frac{3}{4} G_E$.

As $S_{A(E)}n_{A(E)}\propto\sum_{i,j}G_{A(E)}(i,j)$, the proportion that affects the 25\% error rate is related to the sum of all the pixel values of the images. For a full deceiving attack, $G_{\text{all}}=G_E$; thus, only Eve's counts contribute to the 25\% error rate. However, for a partial deceiving attack, Eve's contribution is $4\sum_{i,j}G_{\text{mask}}(i,j)$, where the 4 comes from intercept-and-resend attack. The threshold error rate becomes
\begin{align}\label{pda graphical threshold error}
    e_T = \frac{1}{4}\cdot\frac{4\sum_{i,j}{G_{\text{mask}}(i,j)}}{\sum_{i,j}{G_{\text{all}}(i,j)}} = \frac{\sum_{i,j}{G_{\text{mask}}(i,j)}}{\sum_{i,j}{G_{\text{all}}(i,j)}}.
\end{align}
Thus, we can obtain the error threshold from experimental results since Eq.~\ref{pda errorT by counts} and Eq.~\ref{pda graphical threshold error} are equivalent. Note that this error threshold is calculated under the assumption of an intercept-and-resend attack, which is expected to be the optimal attack strategy, while the error rate in Eq.~\ref{polE} is directly obtained from the coincidence counts.

Let us consider the error values under an ideal intercept-and-resend attack. Since measurement basis choice is random and the state is maximally entangled, $n_{A,X_i}=p(n_{A,X_i}|n_{A}) n_{A}= n_{A}/4$, where $n_{A,X_i}$ denotes an $X_i$-polarized coincidence rate of Alice's photon without a target. A coincidence rate of Eve's photon in $X_i$-polarization is $n_{E,X_i}=p(n_{A,X_i}|n_{A})n_{E}(1/2+1/4)=3n_{E}/16$. The first term, $1/2$, comes from the case where Eve chooses the same basis with Alice's one, and the second, $1/4$, does the case where Eve chooses the wrong basis but the state is projected onto the correct state. Error coincidence came from Eve is $n_{E,e}=p(n_{A,X_i}|n_{A})n_{E}/4=n_{E}/16$, where the $4$ in the denominator comes from the probability that the state is projected onto the error state, which can happen only when Eve chooses the wrong basis. From Eq.~\ref{polE}, the error rate with target information becomes
\begin{align}
	\begin{split}
		e_{X_{i}}&=\frac{S_{E}n_{E,e}}{S_{A}n_{A,X_{i}}+S_{E}(n_{E,X_{i}}+n_{E,e})}\\
		&=\frac{1}{4}\frac{S_{E}n_{E}}{S_{A}n_{A}+S_{E}n_{E}},
	\end{split}
\end{align}
which is equivalent to Eq.~\ref{pda errorT by counts}. Thus, in the context of an intercept-and-resend attack, in ideal, the two values are the same. On the other hand, if Eve performs a different attack which is not optimal, the difference between the error threshold and the error rate would become apparent. For example, if Eve randomly chooses one polarization among the four, the coincidence rate of Eve's photons becomes $n_{E,X_i}=n_{E,e}=n_{E}/8$. Thus, the error rate becomes
\begin{align}
	e_{X_{i}}=\frac{1}{2}\frac{S_{E}n_{E}}{S_{A}n_{A}+S_{E}n_{E}},
\end{align}
which is always greater than the error threshold if $S_{E}n_{E}\neq 0$.

When the error rate is greater than or equal to the partial deceiving attack threshold calculated by Eq.~\ref{pda graphical threshold error}, Eve's partial deceiving attack can be noticed. Extraction of trustworthy information from the obtained images can be performed based on image relations as,
\begin{align}\label{trustworthy}
    G_A(i,j) = G_\text{cor}(i,j) - 3 G_\text{mask}(i,j).
\end{align}

Note that enhanced security of QS-SPI is based on the simultaneous procedure of imaging and security analysis. If the two were done sequentially, coincidence counts for imaging are irrelevant to security checking; thus, neither detection of a partial deceiving attack nor trustworthy image reconstruction is possible.

\begin{figure*}[t!]
	\centering\includegraphics[width=0.95\textwidth]{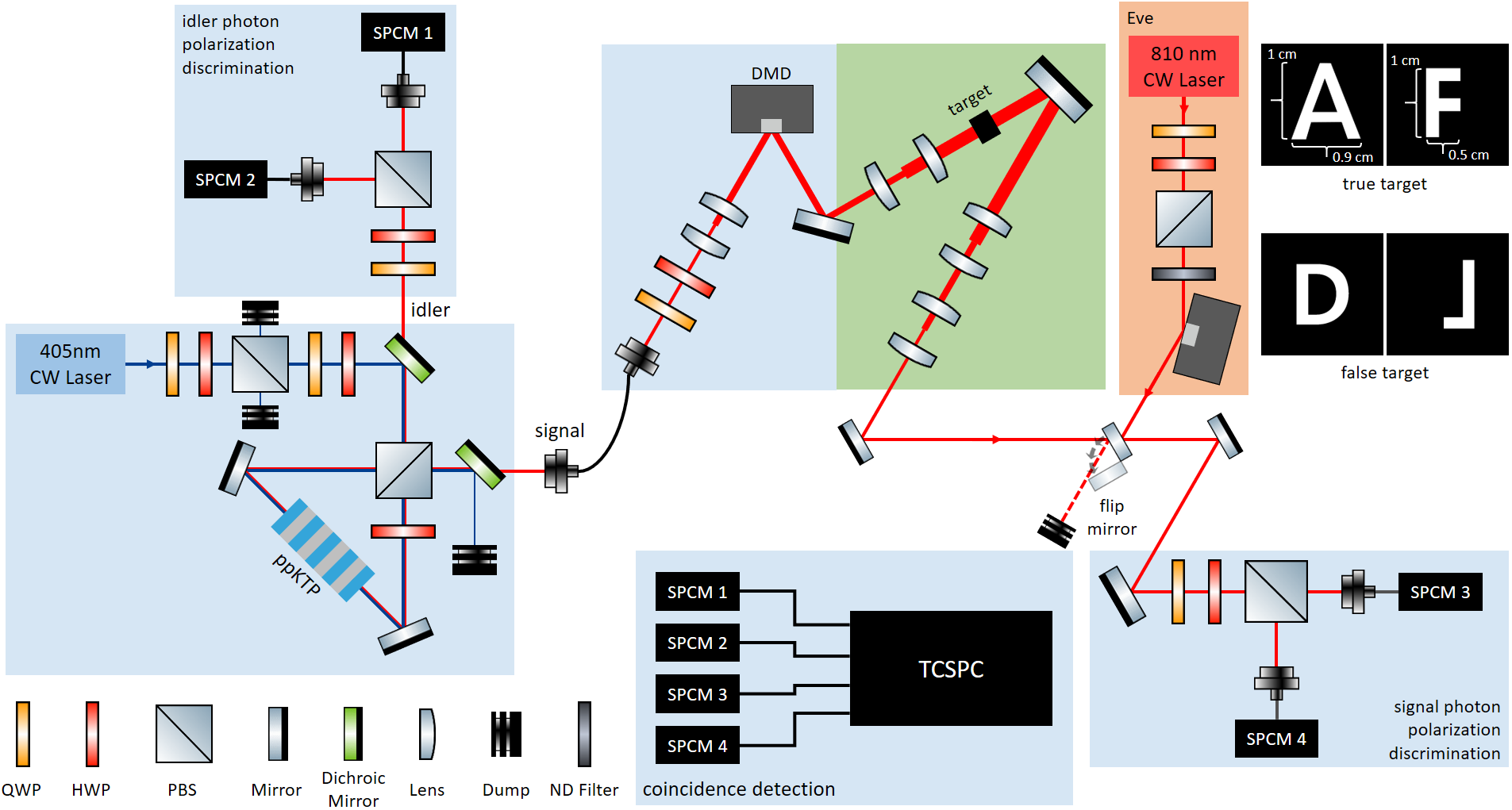}%
	\caption{Experimental setups of our QS-SPI. A polarization-entangled photon pair is generated by the Sagnac interferometer with a ppKTP crystal. Polarization of the idler photon is directly detected by SPCMs. The signal photon is reflected by the DMD with a post-selection of its position and sent to a true target. After interaction with the target, the photon is counted by SPCMs in selected polarization. A full deceiving attack is demonstrated by blocking Alice's signal with a flip mirror and sending Eve's laser beam in a polarization-controlled and intensity-modulated manner. Accidental coincidence counts occur by Eve's light which lead to the formation of a false image. A partial deceiving attack is realized by acquiring photon counts twice: one from a true signal and the other from a false signal. PBS: polarizing beam splitter; QWP (HWP): quarter (half) wave plate; ND filter: neutral density filter.}\label{exp}
\end{figure*}

In summary, the security analysis procedure of QS-SPI is conducted as follows:
\begin{enumerate}
\item If the error rate is greater than or equal to 25\%, the imaging system is under a full deceiving attack. All images should be discarded.
\item If the error rate is smaller than 25\%, Alice calculates $e_T$ based on Eq.~\ref{pda graphical threshold error} to check for a possible partial deceiving attack.
\item If the error rate is smaller than $e_T$, the obtained images are trustworthy. Otherwise, only the trustworthy image reconstructed by Eq.~\ref{trustworthy} is credible.
\end{enumerate}

\section{Proof-of-Principle demonstration}\label{Sec3}
\subsection{QS-SPI setup}

Fig.~\ref{exp} shows the setups for a proof-of-principle demonstration of QS-SPI. A polarization-entangled state is generated from a Sagnac interferometer with a periodically poled potassium titanyl phosphate (ppKTP) crystal \cite{Kim2006}. The crystal is pumped by a 405 nm continuous wave (CW) laser, generating 810 nm polarization-entangled photon pairs via the type-II SPDC process. The initial state generated from the Sagnac interferometer is $\ket{\Psi^{+}}=\frac{1}{\sqrt{2}}\left(\ket{H,V}_{SI}+\ket{V,H}_{SI}\right)$. Therefore, to make the state be $\ket{\Phi^{+}}$, additional phase shifts on the idler mode are given. The state is prepared with $98.6$\% fidelity, verified via quantum state tomography \cite{QST}.

After generation, the idler mode is detected by SPCMs (Excelitas Technologies, SPCM-780-13-FC) in selected polarization. SPCMs are connected to a time-correlated single photon counting (TCSPC) module to record detection time and polarization. The signal photon is sent to the DMD (Vialux GmbH, DLP650LNIR) and post-selected according to a displayed pattern. In our setup, we used Hadamard patterns with a $32 \times 32$ resolution; thus, the total number of shots required for one image is 2048.

The spatially post-selected signal photons interact with the target, alphabet letter ``A'' or ``F'', and are detected by SPCMs. ``A'' is used for demonstrating all attack methods while ``F'' is used for a partial deceiving attack only. TCSPC analyzes detection from the four SPCMs to give the coincidence counts of one signal and one idler.

In the setup, the power of the pump laser for generation of the entangled photon pairs was 5 mW. Single count rates of the signal and idler without a target were $6\times 10^{3}$ cps and $8\times 10^{4}$ cps, respectively. We set the coincidence window as 650 ps, and the coincidence count rate of a signal and an idler in the same polarization was 300 cps. In imaging, the photon acquisition time for one Hadamard pattern was 3.5 seconds.

\subsection{Demonstration of Eve's attack}

\begin{figure*}[t!]
	\centering\includegraphics[width=0.8\textwidth]{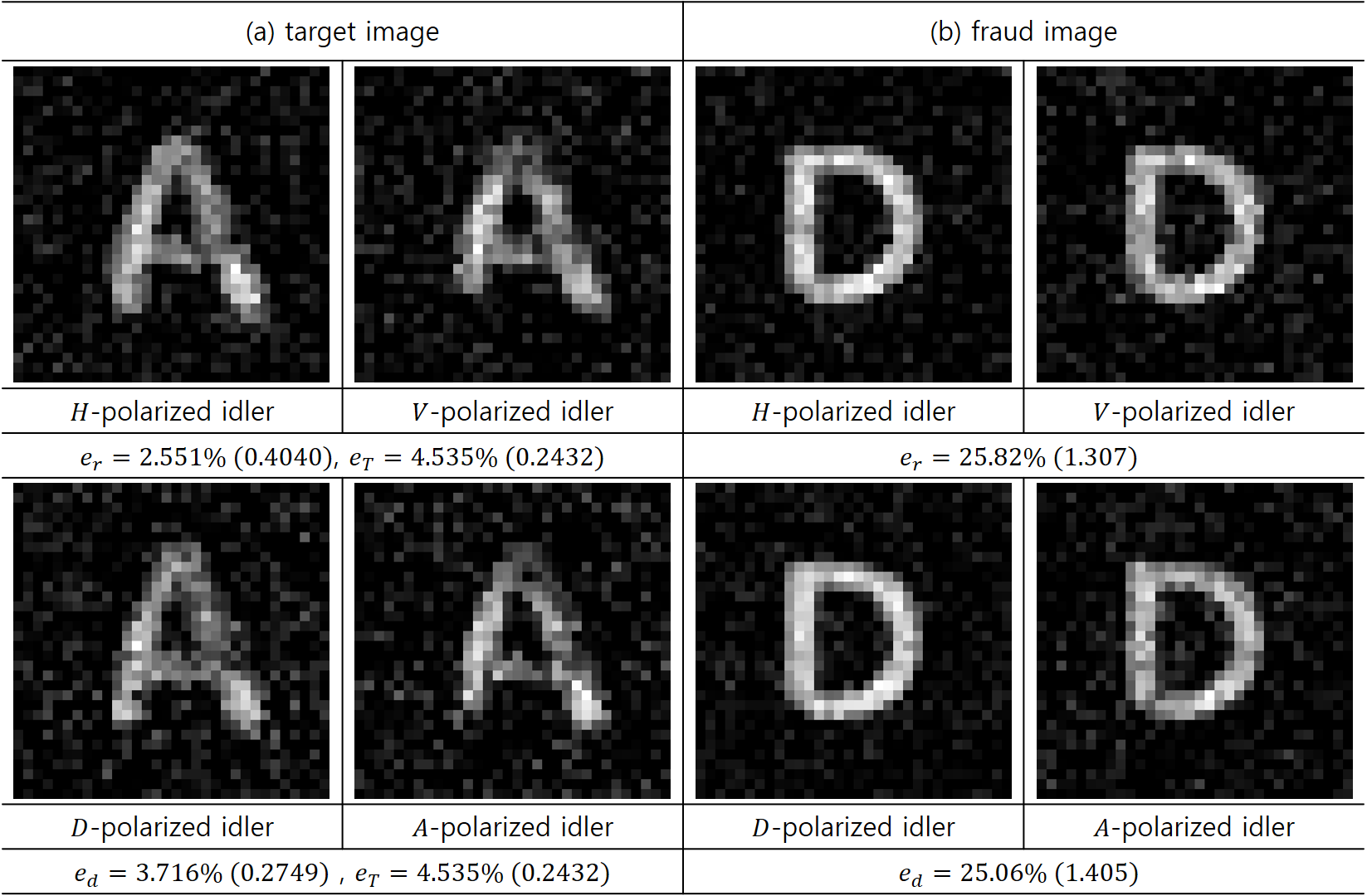}%
	\caption{Images obtained by QS-SPI without any attack (left) or under a full deceiving attack (right). Imaging is performed 5 times and the average error rate in each basis with a standard deviation in parenthesis are shown below the corresponding images, where $e_r$ ($e_d$) denotes the error rate in the rectilinear (diagonal) basis. (a) When there is no attack, all error rates are suppressed below a partial deceiving attack threshold of 4.5\%. (b) Under the attack, error rates close to 25\% are obtained, which is the expected error rate under an ideal intercept-and-resend attack. As the error rates are over 25\%, a full deceiving attack exists, thus all images should be discarded.}\label{result_pol}
\end{figure*}

As previously described, Eve's intercept-and-resend attack exploits an on-demand single-photon source to make a generated photon enter within the coincidence window. However, since the implementation is not feasible with current technologies, we simulated an intercept-and-resend attack with implementable devices, and in the post-process.

In QS-SPI, image reconstruction relies on coincidence counts, and Eve can manipulate the system by controlling these counts. In our proposed deceiving attack, Eve illuminates Alice's receiver with an 810 nm CW laser, inducing accidental coincidence counts. To simplify the attack, we fix the polarization of the illumination laser to $H'=H$ or $D'=D$, instead of controlling the polarization based on the measured information. Ideally, no erroneous coincidence counts occur when Eve chooses the same polarization basis as Alice. For instance, if Alice's signal is in $H$-polarization and Eve chooses the rectilinear basis, Eve sends a false signal in $H$-polarization. However, if Eve selects the diagonal basis, $50\%$ of the counts are recorded as erroneous coincidence counts. To accurately demonstrate the intercept-and-resend attack, we selected coincidence counts for imaging and error rate calculation according to the following criteria:
\begin{enumerate}
\item If Alice and Eve select identical bases, coincidence counts where the polarization of the idler is matched with the false signal are selected, and the others are discarded.
\item If different bases are chosen, all polarization combinations of coincidence counts are selected.
\end{enumerate}
However, for case 1, we ignore Eve's attack in different polarization. For example, we ignore $V$ polarization errors in the rectilinear basis while considering errors in the diagonal basis. To account for this, we consider the possibility of random basis choice for $V$ and $A$ polarized idlers.

In a full deceiving attack, Alice's signal is blocked, and only accidental coincidence counts are detected. To deceive Alice's setup, the accidental coincidence count rate must be similar to the coincidence count rate of the entangled photon pairs. To achieve this condition, the power of Eve's laser is determined as follows. The detection probabilities per window on the signal and idler mode SPCMs are $N_{E}\tau$ and $N_{I}\tau$, respectively, where $N_{E}$ is the photon rate of Eve's laser and $\tau$ is a coincidence window. In this case, the coincidence probability in the coincidence window is given by the product of the single probabilities, $N_{E}N_{I}\tau^{2}$. Then, the accidental coincidence count rate $n_{acc}$ can be calculated by the following equation:
\begin{align}
	n_{acc}=N_{I}N_{E}\tau.
\end{align}
To make $n_{acc}\sim 300$ cps with $N_{I}=8\times 10^{4}$ cps and $\tau = 650$ ps, $N_{E}\sim 5.8\times 10^{6}$ cps is obtained, which is 1000 times greater than the original signal photon count rate without a target. This photon rate corresponds to approximately 1.41 pW for an 810 nm CW laser.

Intensity modulation of Eve's laser, which is necessary to deceive SPI, is performed by using another DMD. Eve's DMD displays the overlapped patterns of Alice's Hadamard patterns and a fraud image. After imaging under a full deceiving attack, a fraud image, the alphabet letter ``D'', is constructed from the accidental coincidence counts.

In a partial deceiving attack, Alice detects both true and false signals with their intensity ratio being controlled. Photon acquisition is performed twice for each imaging pattern. One acquisition is for the true signal and the other is for the false signal, controlled by the flip mirror. The polarization of the false signal is controlled in the same manner as in the full deceiving attack. In this attack, we first set the true target as the alphabet letter ``F''. To disguise the true target profile, the false signal contains target information as the left-and-right inverse of the alphabet letter ``L''. Combining the two signals, a deceiving image is formed: the digital number ``8''. Additionally, we demonstrate the attack with the targets used in the full deceiving attack, namely the alphabet letters ``A'' and ``D''. In this case, the two images overlap, and the true target information is covered by the fraudulent image. 
For the two possible scenarios, we demonstrate trustworthy image reconstruction based on the obtained images.

\begin{figure*}[t!]
	\centering\includegraphics[width=0.8\textwidth]{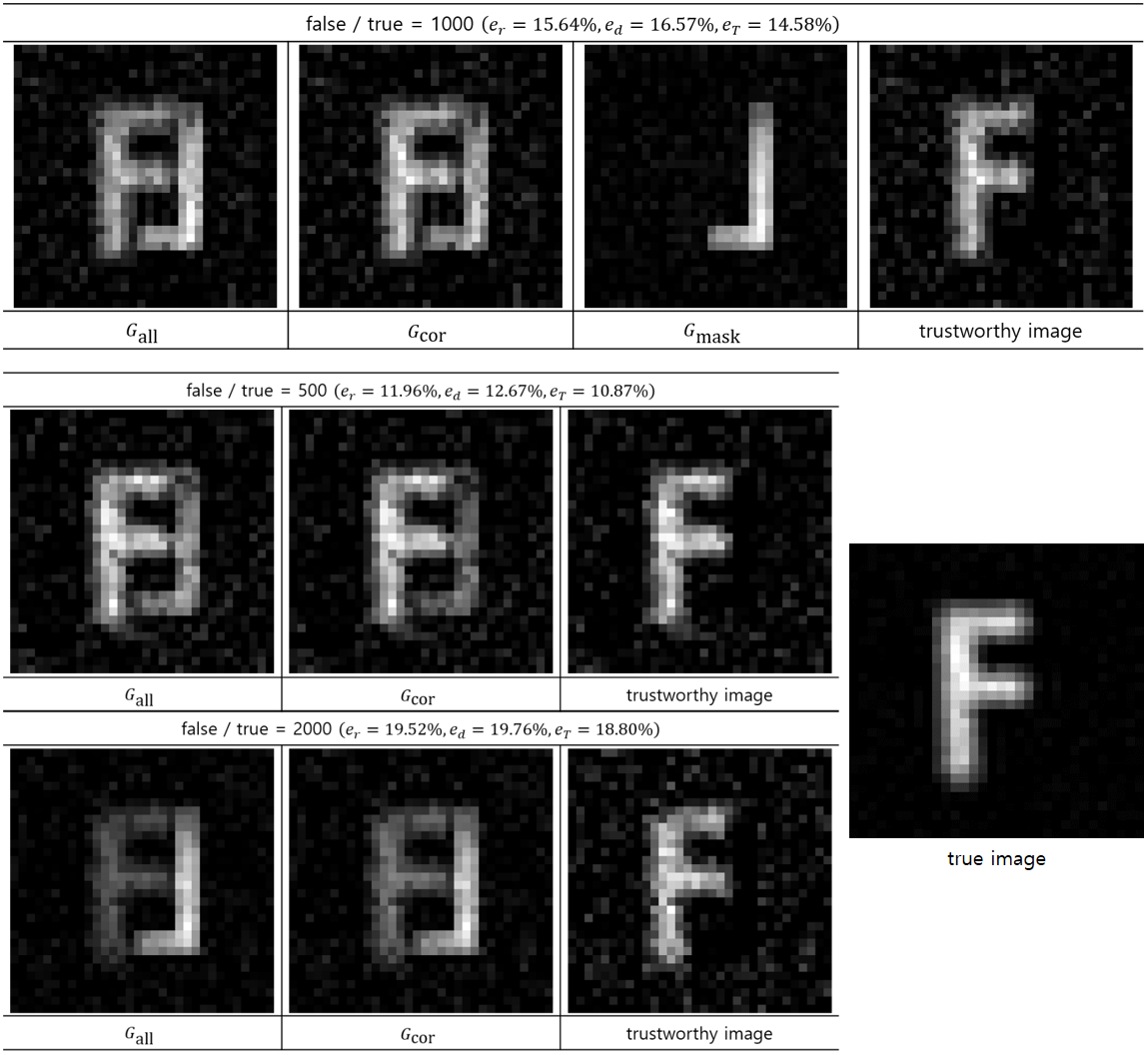}%
	\caption{Images obtained by QS-SPI under a partial deceiving attack where the true target profile has no overlap with the false target profile. A false signal to the true signal intensity ratio of 500, 1000, and 2000 are demonstrated. When the ratio is 1000, a contrast of true and false target images is balanced to form a deceiving image: the digital number ``8''. As either $e_r$ or $e_d$ is beyond $e_T$ but below 25\%, a partial deceiving attack exists. Reconstructed trustworthy images are similar to the true target image. The stronger the noise intensity, the more background noise exists in the trustworthy image due to suppression of the true signal by a strong false signal.}\label{result_pda F8}
\end{figure*}

\begin{figure*}[t!]
	\centering\includegraphics[width=0.8\textwidth]{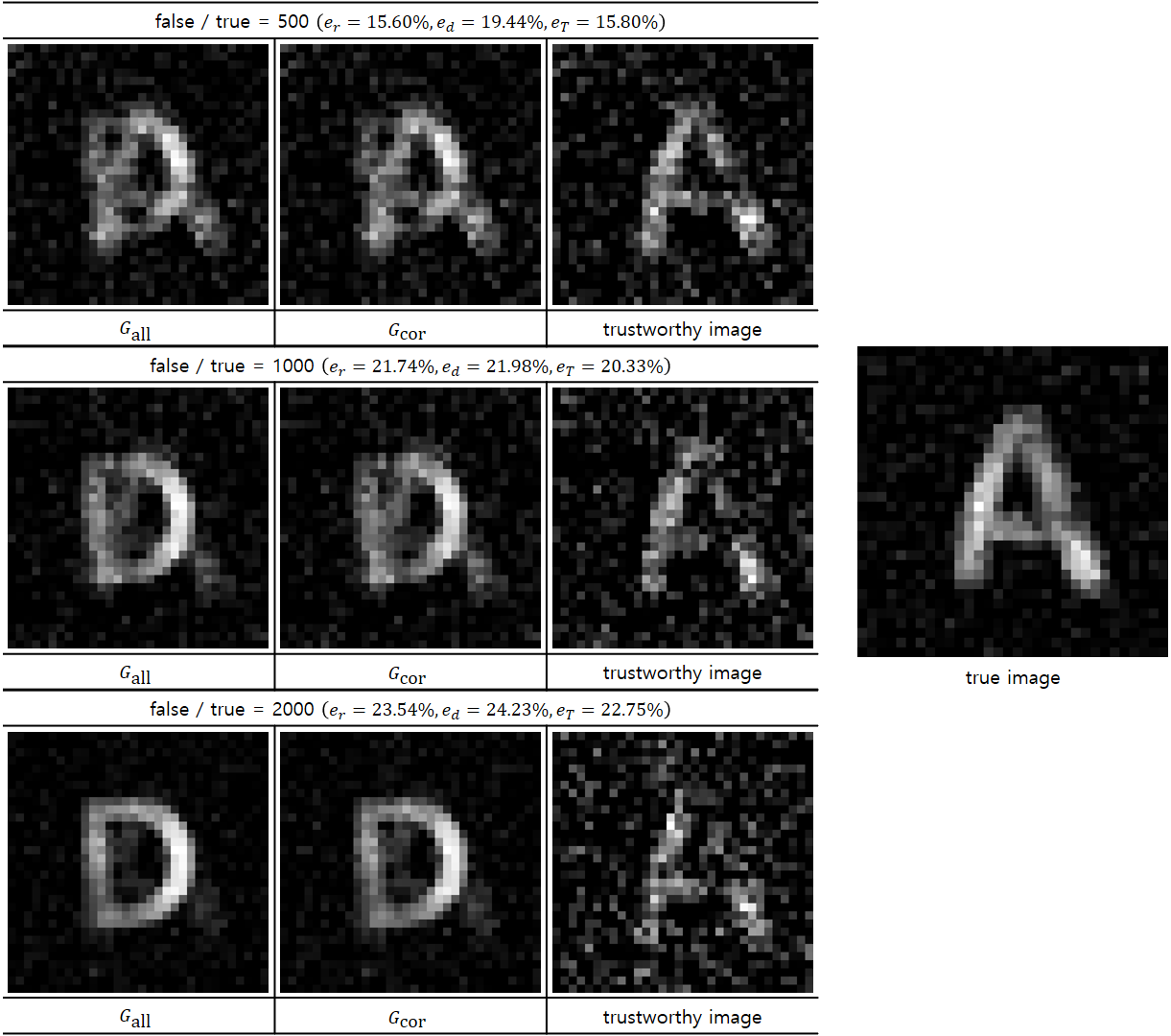}%
	\caption{Demonstration of a partial deceiving attack where a true target profile has overlap with a false target profile. As either $e_r$ or $e_d$ is beyond $e_T$ but below 25\%, a partial deceiving attack exists. From the obtained images, trustworthy images are reconstructed. The image quality of the overlapped region in the trustworthy images worsens as the intensity of the false signal gets stronger. This is due to suppression of the true signal by a strong false signal, which also causes background fluctuation to be stronger.}\label{result_pda AD}
\end{figure*}

\subsection{Results}

Fig.~\ref{result_pol} shows the images obtained without an attack from Eve and under a full deceiving attack. The error rate in each basis is marked below the corresponding images, where the error rate in rectilinear (diagonal) basis is $e_r$ ($e_d$). Due to experimental defects, some pixels in the obtained images were negative; thus, we transitioned the negatives to 0. The images shown are normalized, i.e., all pixels are divided by the maximum pixel value and multiplied by 255.

In the proof-of-principle experiments, all data was obtained five times, and the average of the five error rates is marked with the standard deviation in parenthesis. When there is no attack, error rates lower than 25\% were obtained. To test the possibility of a partial deceiving attack, the threshold error rate was calculated to be 4.54\%. As the error rates are lower than the threshold, the security of the obtained images is guaranteed. Under Eve's full deceiving attack, error rates close to 25\% were obtained as expected from the error analysis under the intercept-and-resend attack. As the error rates are greater than 25\%, Eve's full deceiving attack exists, indicating that the images are fake and should be discarded.

The results of the QS-SPI under a partial deceiving attack are presented in Fig.~\ref{result_pda F8} and Fig.~\ref{result_pda AD}. We varied the false signal intensity to the true signal intensity ratio and demonstrated cases where the ratio is 500, 1000, and 2000. The former is the case where a true target profile has no overlap with a false target profile, but the two are combined to produce a deceiving image. When the ratio is 1000, the contrast of true and false target images is balanced to form a deceiving image: the digital number ``8''. The latter is the case where true and false images have overlapped. True and false targets are the alphabet letters ``A'' and ``D'', respectively, and the overlap deteriorates the true target information.

In both cases, all error rates are below 25\%. However, as either $e_r$ or $e_d$ is greater than the partial deceiving attack threshold $e_T$, a partial deceiving attack exists. We reconstructed trustworthy images using Eq.~\ref{trustworthy} and compared them with the true target image acquired by only the true signals. We observed that $G_\text{mask}$ has the spatial profile of the false target. By subtracting three times of $G_\text{mask}$ from $G_\text{cor}$, we obtained the image of the true target profile. When the error rate is much smaller than 25\%, which is the threshold of a full deceiving attack, the trustworthy images are well recovered (nearly identical to the true images). However, as the error rate approaches 25\%, the background noise increases due to the suppression of the true signal in comparison with the intensity of the false signal. When the two image profiles overlap, the suppression causes pixel values in the overlapped region of 3$G_\text{mask}$ to be similar to those in $G_\text{cor}$. Therefore, some information of the true target profile in the overlapped region may be erased during the trustworthy image reconstruction process, which is why the image quality in the overlapped region rapidly degrades.

\begin{figure}[t!]
	\centering\includegraphics[width=0.4\textwidth]{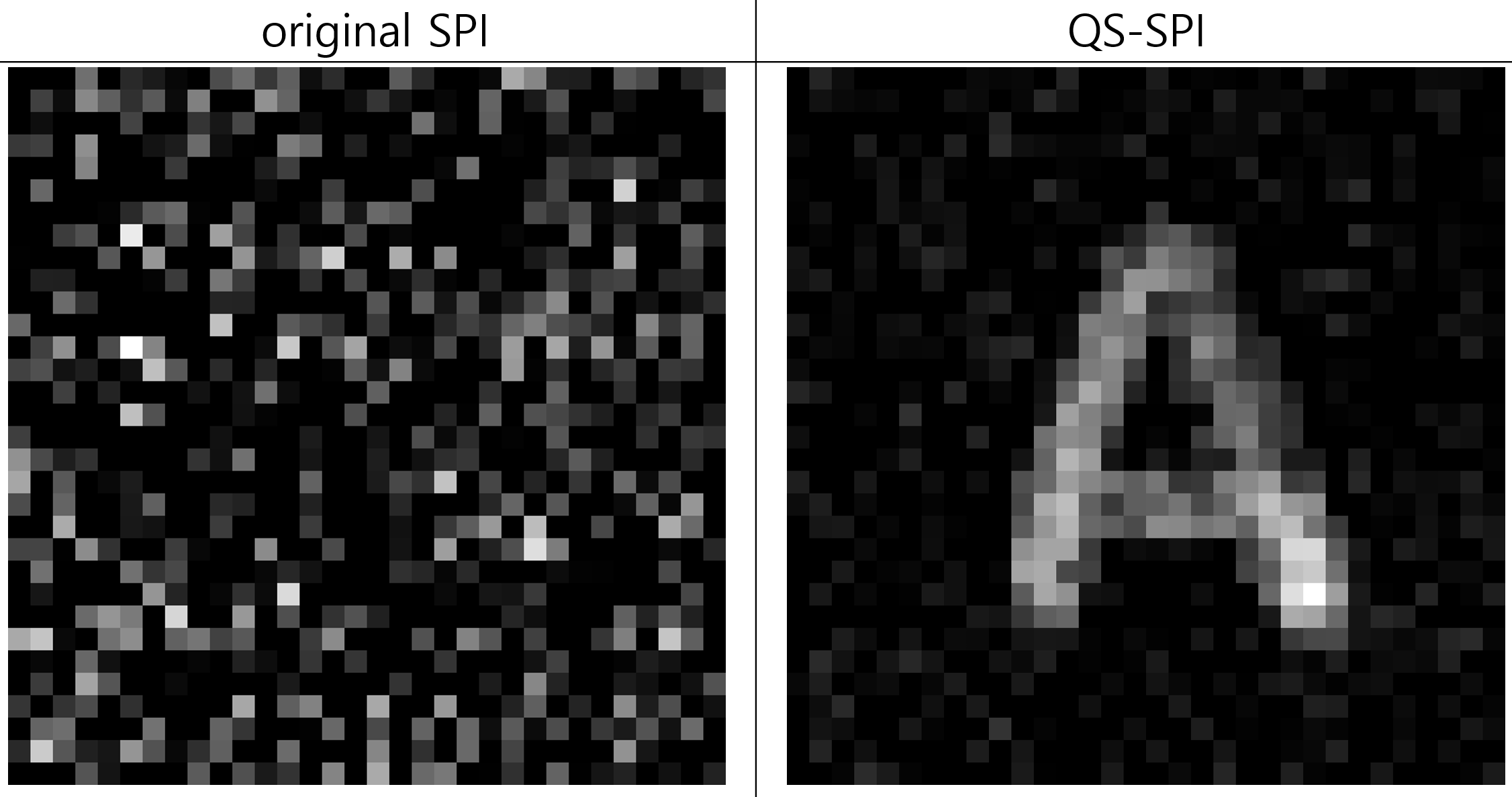}%
	\caption{Images constructed under an imaging disrupting attack by the original SPI (left) and QS-SPI (right). The attack is realized by illuminating Eve's laser to Alice's signal, where the laser power was 1000 times greater than Alice's signal. The image obtained by SPI is ruined, while the time-correlation of a photon pair allowed QS-SPI to successfully construct the image.}\label{Dattack}%
\end{figure}

Lastly, we tested the robustness of QS-SPI against strong noise, and the result is shown in Fig.~\ref{Dattack}. Eve's jamming laser, which has a power 1000 times greater than Alice's signal, illuminates the receiver to disturb the imaging process. In the original SPI, only single counts in the signal mode are exploited to construct an image; thus, the obtained image is ruined by the attack. However, by exploiting the time-correlation between the signal and idler photons, QS-SPI can overcome the attack, allowing the target image to be successfully obtained \cite{Kim2021APL}.

\section{Summary and discussion}\label{Sec4}

In this paper, we proposed QS-SPI, a method that is resilient against possible image deceiving attacks. Our method employs polarization-correlation techniques which enable imaging and security analysis to be performed simultaneously for all detected photon pairs. This results in enhanced security analysis and reconstruction of trustworthy images. We consider two types of deceiving attacks: full deceiving attacks and partial deceiving attacks. A full deceiving attack involves intercepting all true signals and illuminating false signals to construct a fraudulent image. This type of attack can be detected by analyzing statistical errors in polarization-correlation, where the criterion error rate is 25\%. However, a partial deceiving attack involves intercepting a portion of the true signals and mixing them with false signals to create a deceiving image with an error rate below 25\%. To detect this type of attack, we analyze both the polarization-correlation and the obtained images formed by correct and erroneous coincidence counts. When a partial deceiving attack is confirmed, QS-SPI can reconstruct trustworthy information while rejecting false information. We demonstrated the setup of QS-SPI and showed that it is capable of detecting deceiving attacks. We simulated an intercept-and-resend attack, and obtained a theoretical error rate of 25\%. We also showed how QS-SPI can detect partial deceiving attacks and reconstruct trustworthy images from the obtained data, which represents an improvement in security over previously proposed QSI schemes. Finally, we demonstrated the robustness of QS-SPI under strong chaotic noise.

To use QS-SPI as an application, the active basis choice setups and two SPCMs can be replaced with passive ones that consist of a 50:50 beam splitter, phase shifters, and four SPCMs corresponding to the detection of four possible polarization states. In the modified scheme, the photon pairs detected at SPCMs with a mismatched basis should be discarded in the security check. However, a security check based on the Bell inequality \cite{Bell1964}, particularly the Clauser-Horne-Shimony-Holt (CHSH) inequality \cite{Clauser1969}, can be introduced to exploit all the basis combinations to perform security checks \cite{Naik2000}. This approach allows us to construct an image without discarding photon pairs. In the security check, if the polarization statistics violate the Bell inequality, the absence of an attack is guaranteed, and this method provides device-independent security \cite{Barrett2005,Acin2006,Acin2007}.

With small transitions, QS-SPI can be applied to various protocols such as the ghost imaging scheme. In our proposed setup, this can be achieved by removing the DMD and measuring the spatial profile of the idler photon using CCDs or single-pixel detectors in a raster-scanning manner. The role of the DMD is replaced by the measurement of the idler's spatial profile, which provides information about the spatial profile of the signal. When an idler photon is detected, its intensity at a specific pixel position, polarization, and detection time are recorded. By analyzing the intensity-correlation and time-correlation of the signal and idler photons pixel by pixel, an image of the target can be reconstructed with resilience against imaging disrupting attacks. Additionally, by analyzing the polarization-correlation pixel by pixel, the security analysis methodology of QS-SPI can be directly applied. Moreover, QS-SPI is applicable to a quantum-secured optical ranging protocol \cite{Malik2012}. Since our setup exploits the time-correlation of the signal and idler photons, time-of-flight information of a signal photon should be measured. Thus, QS-SPI provides a method to securely acquire a target distance against jamming attacks.

We expect that QS-SPI can advance with matured techniques utilized in quantum secure communication. For instance, six polarization states in three possible mutually unbiased bases (MUBs) can be employed to enhance security \cite{Bruss1998}, or for reference-frame-independent security analysis \cite{Laing2010}. Moreover, various degrees-of-freedom in a single photon can be utilized to exploit high-dimensional quantum states \cite{Cerf2002, Jo2016, Bouchard2018, Jo2019} or hyper-entangled states \cite{Wang2015, JKim2021}. Furthermore, our method can be developed to quantum-secured LiDAR using quantum-correlation-based free-space experimental techniques \cite{DKim2022,Zedda2022}.

\section*{Acknowledgments} This work was supported by a grant to the Defense-Specialized Project funded by the Defense Acquisition Program Administration and Agency for Defense Development.

\bibliography{QSSPI}

\end{document}